\documentstyle[preprint,aps]{revtex}

\title{\vspace*{1pt} \Large \bf
       A new leapfrog integrator of rotational motion. The revised
       angular-momentum approach \\ [6pt]}

\author{\large Igor P. Omelyan \\ [6pt]}

\address{Institute for Condensed Matter Physics,
         National Ukrainian Academy of Sciences, \\ [-2pt]
         1 Svientsitsky st., UA-290011 Lviv, Ukraine.
         E-mail: nep@icmp.lviv.ua \\ [12pt]}

\begin{document}

\maketitle

\vspace{0.4cm}

\begin{abstract}
A new algorithm is introduced to integrate the equations of rotational
motion. The algorithm is derived within a leapfrog framework and the
quantities involved into the integration are mid-step angular momenta
and on-step orientational positions. Contrary to the standard implicit
method by Fincham [Mol. Simul., 8, 165 (1992)], the revised angular
momentum approach presented corresponds completely to the leapfrog idea
on interpolation of dynamical variables without using any extrapolations.
The proposed scheme intrinsically preserves rigid molecular structures and
considerably improves stability properties and energy conservation. As is
demonstrated on the basis of simulations for water, it allows to reproduce
correct results with extra large step sizes of order 5 fs and 10 fs in the
cases of energy- and temperature-conserving dynamics, respectively. We show
also that iterative solutions can be avoided within our implicit scheme
shifting from quaternions to the entire rotation-matrix representation.
\end{abstract}

\vspace{1.25cm}


\section{Introduction}

Computer experiment by the method of molecular dynamics (MD) is intensively
exploited in solving various tasks of chemical physics [1], biochemistry [2]
and biology [3]. Among these are investigations of structure and dynamical
properties of molecular liquids which normally are treated as collections
of rigid bodies. Despite the long prehistory of MD simulation, the
development of efficient and stable algorithms for the integration
of motion for such systems still remains an actual problem.

Usually, molecular movements are simulated using constrained dynamics
[4--7] in which the phase trajectory of each atom is evaluated by Newton's
equations, while the molecular structures are maintained by holonomic
constraints to keep intramolecular bond distances. Although the
atomic-constraint technique can be applied, in principle, to arbitrary
polyatomics regardless of its chemical structure and size, it appears
to be very sophisticated to implement for some particular models. For
example, when there are more than two, three or four interaction sites
per molecule for linear, planar and three-dimensional bodies, bond
lengths and angles cannot be fixed uniquely [5]. Systems of point
molecules with embedded multipoles present additional complexities
too, because then the intermolecular forces cannot easily be decomposed
into direct site-site interactions. The limitation of constrained dynamics
is also caused by the fact that constraint forces are calculated at each
time step of the produced trajectory to balance all other potential forces
in the system. As the number of atoms in each molecule is increased, the
number of constraints raises dramatically, resulting in a decreased speed
of computations. Moreover, to reproduce the rigid molecular structure,
cumbersome systems of nonlinear equations must be solved iteratively. This
can lead to a problem for molecules with light hydrogen atoms or with
linear or planar fragments. In this case, the algorithm converges rather
slowly [8] already at relative small step sizes and, thus, it requires a
considerably portion of the computational time. Recently, it was shown that
a non-iterative calculation of constraint forces is possible [9], but this
is practical only for simple models in which the problem can be reduced to
inversion of a banded matrix [10, 11].

Some of the limitations just mentioned are absent in the molecular approach,
when the displacements of rigid bodies are analyzed in view of translational
and rotational motions. The translational dynamics is defined by motion
of molecular centres of masses, whereas the orientations typically are
expressed in terms of quaternions [12--14] or principal-axis vectors [13].
The straightforward parameterization of orientational degrees of freedom,
Euler angles, is very inefficient for numerical calculations because of
singularities inherent in the description [12, 15, 16]. Multistep
predictor-corrector methods were applied to integrate rotational motion
in early investigations [17--20]. As was soon established, the extra order
obtained in these methods is not relevant, because the forces existing in
a real system are not sufficiently smooth. As a result, high-order schemes
appear to be less accurate at normal step sizes than low-order integrators,
such as Verlet [21], velocity Verlet [22] and leapfrog [23] ones. The last
algorithms are also the most efficient in view of cost measured in terms of
force evaluations. That is why, they are widely used in different approaches,
for instance, in the atomic-constraint technique, to integrate translational
motion. These traditional algorithms were derived, however, assuming that
velocities and forces are coordinate- and velocity-independent, respectively.
In general, the time derivatives of orientational positions may depend not
only on angular velocities but also on these positions themselves resulting
in the explicit velocity-dependence of angular accelerations. Therefore,
additional revisions are necessary to apply the standard integrators to
rotational motion.

In the atomic approach, the problem with the coordinate and velocity
dependencies is circumvented by involving fundamental variables, namely,
the individual Cartesian coordinates of atomic sites. Similarly, this
problem can be solved within the molecular approach choosing appropriate
generalized coordinates in orientational space. Ahlrichs and Brode proposed
a method [24] in which the principal axes of molecules are treated as
pseudo particles and constraint forces are introduced to maintain their
orthonormality. Kol {\em et al.} considered the entire rotation matrix
and the corresponding conjugate momentum as dynamical variables [25]. The
rotation matrices can be evaluated within the usual Verlet or leapfrog
frameworks, using either recursive [24] or iterative [25] procedures,
respectively. The recursive method behaves relatively poor with respect
to long-term stability of energy, whereas the iterative procedure requires
again, as in the case of constrained dynamics, to find solutions for systems
of highly nonlinear equations. In general, the convergence of iterations
is not guaranteed and looping becomes possible even at not very large step
times. Examples for not so well behaved cases are models with almost linear
or planar molecules, when the diagonal mass matrices are hard to numerical
inversion since they have one or two elements which are very close to
zero. The extension of the atomic and pseudo-particle approaches to
temperature-conserving dynamics is also a difficult problem, given that
the rigid-reactions and temperature-constraint forces are coupled between
themselves in a very complicated manner.

A viable alternative to integrate the rigid-body motion has been provided
by explicit and implicit angular-momentum algorithms of Fincham [26--27].
This was the first attempt to adopt the leapfrog framework to rotational
motion in its purely classical treatment. The chief advantage of these
rotational leapfrog algorithms is the possibility to perform thermostatted
simulations. However, even in the case of a more stable implicit algorithm,
the total energy fluctuations in energy-conserving simulations are too
big with respect to those identified in the atomic-constraint technique.
Moreover, despite the fact that no constraint forces are necessary in the
rotation dynamics, the rigidness of molecules is not satisfied automatically,
because the equations of motion are not solved exactly. Usually, the
artificial rescaling method [19, 27] is used to preserve the unit norm
of quaternions and, as a consequence, to ensure the molecular rigidity.
Recently [28], it has been shown that the crude renormalization can be
replaced by a more rigorous procedure introducing so-called numerical
constraints. As a result, quaternion [28] and principal-axis [29]
algorithms were devised within the velocity Verlet framework. It was
demonstrated [29, 30] that these algorithms conserve the total energy
better than the implicit leapfrog integrator [27], but worse with respect
to the atomic-constraint method, especially in the case of long-duration
simulations with large step sizes.

Quite recently, to improve the stability, a new angular-velocity leapfrog
algorithm for rigid-body simulations has been introduced [30]. The automatic
preservation of rigid structures and good stability properties can be related
to its main advantages. But a common drawback, existing in all long-term
stable integrators on rigid polyatomics, still remained here, namely, the
necessity to solve by iteration the systems of nonlinear equations. Although
such equations are much simpler than those arising in the atomic and
pseudo-particle approaches, the iterative solution should be considered
as a negative feature. Moreover, since the nonlinear equations are with
respect to velocities, it is not so simple matter to extend the
angular-velocity algorithm to a thermostatted version.

This study presents a modified formulation of the angular-momentum approach
within the leapfrog framework. Unlike the standard approach by Fincham [27],
the new formulation is based on more natural interpolations of dynamical
variables and it uses no extrapolation. The algorithm derived appears to
be free of all the drawbacks inherent in previous descriptions. It
can easily be implemented to arbitrary rigid bodies and applied to
temperature-conserving dynamics. The integrator exhibits an excellent
energy conservation, intrinsically reproduces rigid structures and
allows to avoid any iterative procedures at all.

\section{Basic equations of motion}

Let us consider a system of $N$ interacting rigid bodies. According to the
classical approach, any movements of a body can be presented as the sum of
two motions, namely, a translational displacement of the centre of mass and
a rotation about this centre. The translational displacements in the system
are expressed in terms of the centre-of-mass positions ${\bf r}_i$ and
velocities ${\bf v}_i$, where $i=1,\ldots,N$, given in a space-fixed
laboratory frame. The time evolution of such quantities can described
by writing Newton's law in the form of two per particle three-dimensional
differential equations of first order,
\begin{eqnarray}
m_i \frac{{\rm d} {\bf v}_i}{{\rm d} t}&=&{\bf f}_i \nonumber \, ,
\\ [-6pt] \\ [-6pt]
\frac{{\rm d} {\bf r}_i}{{\rm d} t}&=&{\bf v}_i \, , \nonumber
\end{eqnarray}
where ${\bf f}_i$ is the total force acting on body $i$ due to the
interactions with all the rest of particles and $m_i$ denotes the mass
of the body.

\subsection{Different forms of the equations for rotational motion}

To determine the rotational motion, one needs to use frames attached to
each body together with the laboratory system of coordinates. It is more
convenient for further consideration to direct the body-fixed-frame axes
along the principal axes of the particle, which pass through its centre of
mass. Then the matrix ${\bf J}_i$ of moments of inertia will be diagonal
and time-independent in the body-fixed frame. We will use the convention
that small letters stand for the representation of variables in the fixed
laboratory frame, whereas their counterparts in the body frame will be
designated by capital letters. The transitions ${\bf E}={\bf A}_i {\bf e}$
and ${\bf e}={\bf A}_i^{-1} {\bf E}$ between these both representations
of vectors ${\bf e}$ and ${\bf E}$ in the laboratory and body frames,
respectively, can be defined by the $3 \times 3$ time-dependent rotation
matrix ${\bf A}_i(t)$. Such a matrix must satisfy the orthogonormality
condition ${\bf A}_i^+ {\bf A}_i={\bf I}={\bf A}_i {\bf A}_i^+$, or in
other words ${\bf A}_i^{-1}={\bf A}_i^+$, to ensure the invariance ${\bf
E}^+ {\bf E}={\bf e}^+ {\bf e}$ of quadratic norms for vectors ${\bf e}$
and ${\bf E}$. In our notations ${\bf A}^{-1}$ and ${\bf A}^+$ are the
matrices inversed and transposed to ${\bf A}$, correspondingly, and ${\bf
I}$ denotes the unit matrix.

Let ${\mbox{\boldmath $\Delta$}}_i$ be an arbitrary vector fixed in the
body. By definition, such a vector does not change in time in the body-fixed
frame, ${\rm d} {\mbox{\boldmath $\Delta$}}_i/{\rm d} t=0$. The angular
velocity ${\mbox{\boldmath $\omega$}}_i$ is introduced differentiating its
counterpart ${\mbox{\boldmath $\delta$}}_i(t)={\bf A}_i^+(t) {\mbox{\boldmath
$\Delta$}}_i$ in the laboratory frame over time, ${\rm d} {\mbox{\boldmath
$\delta$}}_i/{\rm d} t={\mbox{\boldmath $\omega$}}_i {\mbox{\boldmath $\times
\delta$}}_i$. Then, using the equality ${\mbox{\boldmath $\omega$}}_i {\mbox
{\boldmath $\times \delta$}_i}={\bf W}^+({\mbox{\boldmath $\omega$}}_i)
{\mbox{\boldmath $\delta$}_i}$ and the orthonormality of ${\bf A}_i$, the
rate of change in time of the orientational matrix can be expressed in terms
of either laboratory ${\mbox{\boldmath $\omega$}}_i$ or principal ${\mbox
{\boldmath $\Omega$}}_i={\bf A}_i {\mbox{\boldmath $\omega$}}_i$ angular
velocity as
\begin{equation}
\frac{{\rm d} {\bf A}_i}{{\rm d} t}=
{\bf A}_i {\bf W}({\mbox{\boldmath $\omega$}}_i)=
{\bf W}({\bf \Omega}_i) {\bf A}_i \, ,
\end{equation}
where
\vspace{-6pt}
\begin{equation}
{\bf W}({\bf \Omega}_i)=\left(
\begin{array}{ccc}
0 & \Omega_Z^i & -\Omega_Y^i \\
-\Omega_Z^i & 0 & \Omega_X^i \\
\Omega_Y^i & -\Omega_X^i & 0
\end{array}
\right)
\end{equation}
is a skewsymmetric matrix, i.e., ${\bf W}^+({\bf \Omega}_i)=-{\bf W}({\bf
\Omega}_i)$, and $\Omega_X^i$, $\Omega_Y^i$ and $\Omega_Z^i$ are components
of vector ${\bf \Omega}_i$.

From the orthogonormality condition it follows that maximum three independent
parameters are really necessary to describe orientations of a rigid body and
to evaluate the nine elements of the rotation matrix. However, the well-known
parameterization of ${\bf A}_i$ in terms of three Eulerian angles [12] is
unsuitable for numerical calculations because of the singularities. In the
body-vector representation [13, 24, 25, 29], all the elements of the rotation
matrix ${\bf A}_i$ are considered as dynamical variables. These variables
present, in fact, Cartesian coordinates of three principal axes $XYZ$ of the
body in the laboratory frame. The alternative approach applies the
quaternion parameterization [4, 9] of rotation matrices,
\begin{equation}
{\bf A} ({\bf q}_i) =
\left( \begin{array}{ccc}
-\xi_i^2+\eta_i^2-\zeta_i^2+\chi_i^2 & 2(\zeta_i \chi_i - \xi_i \eta_i) &
2 (\eta_i \zeta_i + \xi_i \chi_i) \\
-2(\xi_i \eta_i + \zeta_i \chi_i) & \xi_i^2-\eta_i^2-\zeta_i^2+\chi_i^2 &
2(\eta_i \chi_i - \xi_i \zeta_i) \\
2(\eta_i \zeta_i - \xi_i \chi_i) & -2(\xi_i \zeta_i + \eta_i \chi_i) &
-\xi_i^2-\eta_i^2+\zeta_i^2+\chi_i^2
\end{array}
\right) \, ,
\end{equation}
where ${\bf q}_i \equiv (\xi_i,\eta_i,\zeta_i,\chi_i)^+$ is a vector-column
consisting of four quaternion components. Using the normalization condition
${\bf q}_i^+ {\bf q}_i=\xi_i^2+\eta_i^2+\zeta_i^2+\chi_i^2=1$, which ensures
the orthonormality of ${\bf A}_i(t) \equiv {\bf A} [{\bf q}_i(t)]$, the time
derivatives of quaternions can be cast [13, 27, 30] in the form
\begin{equation}
\frac{{\rm d} {\bf q}_i}{{\rm d}t}
= \displaystyle \frac12 \left(
\begin{array}{cccc}
0&\Omega_Z^i&-\Omega_X^i&-\Omega_Y^i \\
-\Omega_Z^i&0&-\Omega_Y^i&\Omega_X^i\\
\Omega_X^i&\Omega_Y^i&0&\Omega_Z^i  \\
\Omega_Y^i&-\Omega_X^i&-\Omega_Z^i&0
\end{array}
\right)
\left( \begin{array}{c}
\xi_i \\ \eta_i \\ \zeta_i \\ \chi_i
\end{array} \right)
\equiv \displaystyle {\bf Q}({\bf \Omega}_i) {\bf q}_i \, ,
\end{equation}
where ${\bf Q}({\bf \Omega}_i)$ is a skewsymmetric matrix again.

Expressions (2) and (5) are rotation-motion analogues of the second line
of Eq.~(1) in the case of body-vector and quaternion representations,
respectively. They must be complemented by equations defining the time
evolution of angular velocities. The simplest form of these equation is
obtained for the angular momenta ${\bf l}_i={\bf A}_i^+ {\bf L}_i$ of
bodies in the laboratory frame, where ${\bf L}_i={\bf J}_i {\bf \Omega}_i$
are principal angular momenta. The result is
\begin{equation}
\frac{{\rm d} {\bf l}_i}{{\rm d} t}={\bf k}_i \, ,
\end{equation}
where ${\bf k}_i$ is the torque exerted on body $i$ with respect to its
centre of mass. The angular velocities in the body- or space-fixed frames
can easily be reproduced, whenever they are needed, using the relations
${\bf \Omega}_i={\bf J}_i^{-1} {\bf A}_i {\bf l}_i$ and ${\mbox{\boldmath
$\omega$}}_i={\bf A}_i^+ {\bf \Omega}_i={\bf j}_i^{-1} {\bf l}_i$, where
${\bf j}_i={\bf A}_i^+ {\bf J}_i {\bf A}_i$ is the time-dependent matrix of
moments of inertia in the laboratory frame. Another way lies in involving
explicit equations for principal angular velocities. Such equations, known
also as Euler's ones, can be derived substituting ${\bf l}_i={\bf A}_i^+
{\bf J}_i {\bf \Omega}_i$ into Eq.~(6) and using equations of motion (2)
for orientational matrices. As a result, one obtains
\begin{equation}
\frac{{\rm d} {\bf \Omega}_i}{{\rm d} t}= {\bf J}_i^{-1} [
{\bf K}_i + {\bf W}({\bf \Omega}_i) {\bf J}_i {\bf \Omega}_i] \, ,
\end{equation}
where ${\bf K}_i={\bf A}_i {\bf k}_i$ are the principal torques.
Formally replacing the quantities
${\bf \Omega}_i$, ${\bf K}_i$ and ${\bf J}_i$ by ${\mbox{\boldmath
$\omega$}}_i$, ${\bf k}_i$ and ${\bf j}_i$ yields quite similar equations
of motion for angular velocities ${\mbox{\boldmath $\omega$}}_i$ in the
laboratory frame.

It is worth remarking that the body-vector (Eq.~(2)) and quaternion
(Eq.~(5)) representations as well as the angular-momentum (Eq.~(6)) and
angular-velocity (Eq.~(7)) approaches are completely equivalent between
themselves from the mathematical point of view. For numerical evaluations,
the preference must be given to equations which allow to be integrated in
the simplest manner with the greatest precision and the best stability. In
the present study we shall deal with more simple equations of motion (6)
for angular momenta in the laboratory frame rather than with equations (7)
for principal angular velocities. In such a way, difficulties with the
velocity-dependence of angular accelerations are excluded automatically.
Moreover, we shall show that the angular-momentum approach allows to
obviate iterative solutions within a leapfrog framework choosing the
entire-rotation-matrix elements, instead of quaternions, as orientational
variables. Thus, the body-vector representation should be considered as a
more preferable method for such calculations.

\section{The revised angular-momentum approach}

In the case of translational motion, equations (1) can readily be integrated
with the help of the usual [23] leapfrog algorithm:
\begin{eqnarray}
{\bf v}_i(t+{\textstyle \frac{h}{2}})&=&
{\bf v}_i(t-{\textstyle \frac{h}{2}})+
h\,{\bf f}_i(t)/m + {\cal O}(h^3) \, , \nonumber \\ [-12pt] \\ [-12pt]
{\bf r}_i(t+h)&=&{\bf r}_i(t)+h\,{\bf v}_i(t+{\textstyle \frac{h}{2}})
+ {\cal O}(h^3) \nonumber \, ,
\end{eqnarray}
where $h$ is the time increment, and forces ${\bf f}_i(t)$ are evaluated
using known spatial coordinates ${\bf r}_i(t)$ and ${\bf A}_i(t)$. The
truncation local errors, appearing during such an integration, are of order
$h^3$ in both coordinates and velocities. If an estimator of ${\bf v}_i(t)$
is required, for example to evaluate the total energy, the usual choice is
\begin{equation}
{\bf v}_i(t)=\frac12 \Big[{\bf v}_i(t-{\textstyle \frac{h}{2}})+
{\bf v}_i(t+{\textstyle \frac{h}{2}}) \Big] + {\cal O}(h^2) \, ,
\end{equation}
where interpolation uncertainties ${\cal O}(h^2)$ are in the self-consistency
with the second order of global errors (one order lower than that for local
errors) of the leapfrog integrator (8).

\subsection{Standard rotational leapfrog algorithm}

For the rotational motion the time derivatives of orientational positions
(Eqs.~(2) and (5)) depend not only on angular velocities but also on these
positions themselves. This difficulty cannot be handled with a simple
leapfrog scheme in which the positions and velocities are known at different
times. Relatively recently, Fincham [27] has proposed a solution to the
problem by introducing an implicit leapfrog-like algorithm. His method
can briefly be described as follows.

First, quite analogously to the case of translational-velocity evaluations
(fist line of Eq.~(8)), angular-momentum equations (6) are integrated as
\begin{equation}
{\bf l}_i(t+{\textstyle \frac{h}{2}}) =
{\bf l}_i(t-{\textstyle \frac{h}{2}})+
h\,{\bf k}_i(t) + {\cal O}(h^3) \, .
\end{equation}
At this stage the principal angular velocities ${\bf \Omega}_i(t)$ can be
calculated using the relation
\vspace{-8pt}
\begin{equation}
{\bf \Omega}_i(t)={\bf J}_i^{-1} {\bf A}_i(t) {\bf l}_i(t)
\end{equation}
and the propagation
\vspace{2pt}
\begin{equation}
{\bf l}_i(t) = \frac12
\Big[ {\bf l}_i(t-{\textstyle \frac{h}{2}}) +
{\bf l}_i(t+{\textstyle \frac{h}{2}}) \Big]=
{\bf l}_i(t-{\textstyle \frac{h}{2}}) + {\textstyle \frac{h}{2}}
{\bf k}_i(t) + {\cal O}(h^2)
\end{equation}
of angular momenta to on-step level of time.

Further, according to the leapfrog framework, the evaluation of orientational
coordinates must be performed as
\begin{equation}
{\bf S}_i(t+h)={\bf S}_i(t)+h
{\bf {\dot S}}_i(t+{\textstyle \frac{h}{2}}) + {\cal O}(h^3) \, ,
\end{equation}
where ${\bf {\dot S}}_i \equiv {\rm d} {\bf S}_i/{\rm d} t={\bf H}({\bf
\Omega}_i) {\bf S}_i$, and either ${\bf S}_i \equiv {\bf A}_i$ and ${\bf
H} \equiv {\bf W}$ or ${\bf S}_i \equiv {\bf q}_i$ and ${\bf H} \equiv
{\bf Q}$ in the case of either entire-matrix or quaternion space,
respectively. Note that in the quaternion representation the orientational
matrices ${\bf A}_i(t) \equiv {\bf A}_i[{\bf q}_i(t)]$ appear implicitly,
and they are computed via relation (4) using quaternion values ${\bf
q}_i(t)$. As far as the quantities ${\bf S}_i$ and ${\bf \Omega}_i$ are
not known at mid-step level $t+{\textstyle \frac{h}{2}}$, it was assumed
to propagate the time derivatives of ${\bf S}_i$ by means of the relation
${\bf {\dot S}}_i(t+{\textstyle \frac{h}{2}})={\textstyle \frac12} [{\bf
{\dot S}}_i(t)+{\bf {\dot S}}_i(t+h)] + {\cal O}(h^2)$, i.e.,
\vspace{2pt}
\begin{equation}
{\bf {\dot S}}_i(t+{\textstyle \frac{h}{2}})=
\frac12 \Big[
{\bf H}({\bf \Omega}_i(t)) {\bf S}_i(t)+
{\bf H}({\bf {\tilde \Omega}}_i(t+h)) {\bf S}_i(t+h)
\Big] + {\cal O}(h^2) \, ,
\end{equation}
where
\vspace{-8pt}
\begin{equation}
{\bf {\tilde \Omega}}_i(t+h)={\bf J}_i^{-1}
{\bf A}_i(t+h) {\bf {\tilde l}}_i(t+h) \, .
\end{equation}
Propagation (14) requires in its turn the knowledge of advanced angular
momenta ${\bf {\tilde l}}_i(t+h)$ which were predicted by writing
\begin{equation}
{\bf {\tilde l}}_i(t+h) = {\bf l}_i(t+{\textstyle \frac{h}{2}}) +
{\textstyle \frac{h}{2}} {\bf k}_i(t) + {\cal O}(h^2) \, .
\end{equation}

In view of (14) and (15), relation (13) is an implicit system of equations
with respect to elements of ${\bf S}_i(t+h)$, defined through the auxiliary
parameters ${\bf l}_i(t)$ and ${\bf {\tilde l}}_i(t+h)$ which are not stored,
but used to calculate the angular velocities ${\bf \Omega}_i(t)$ and ${\bf
{\tilde \Omega}}_i(t+h)$ in the body frame. The system can be solved by
iteration taking ${\bf S}_i^{(0)}(t+h)={\bf S}_i(t) + h {\bf H}({\bf
\Omega}_i(t)) {\bf S}_i(t)$ as the initial guess.

A thermostatted version is based on interpolations (9) and (12) of on-step
translational velocities and angular momenta. Such interpolations are used
in microcanonical simulations to evaluate the kinetic temperature $T(t)=T(
\{ {\bf v}_i(t),{\bf \Omega}_i(t) \})= \frac{1}{l N k_{\rm B}} \sum_{i=1}^N
[m {{\bf v}_i}^2(t) + \sum_{\alpha}^{X,Y,Z} J_{\alpha \alpha}^i {\Omega_
\alpha^i}^{\!2}(t)]$, where $J_{XX}^i$, $J_{YY}^i$ and $J_{ZZ}^i$ are
nonzero elements of matrix ${\bf J}_i$, $k_{\rm B}$ is the Boltzmann's
constant and $l=6$ denotes the number of degrees of freedom per particle
(for linear bodies $l=5$). This allows to synchronize in time the
temperature with the potential energy $U(t) \equiv U(\{ {\bf r}_i(t),
{\bf S}_i(t) \})$ and, therefore, to calculate the total energy
$E(t)=\frac{l N k_{\rm B}}{2} T(t)+U(t)$ of the system. In the
temperature-conserving dynamics, on-step velocities and angular momenta
are modified as ${\bf v}_i'(t)=\beta(t) {\bf v}_i(t)$ and ${\bf l}_i'(t)=
\beta(t) {\bf l}_i(t)$ using the scaling factor $\beta(t)=\sqrt{T_0/T(t)}$,
where $T_0$ is the required constant temperature [27, 31]. The velocity
integration is completed by
\begin{eqnarray}
{\bf v}_i'(t+{\textstyle \frac{h}{2}})&=&
[2-\beta^{-1}(t)] {\bf v}_i'(t)+{\textstyle \frac{h}{2}} {\bf f}_i(t)/m
\, , \\ [3pt]
{\bf l}_i'(t+{\textstyle \frac{h}{2}})&=&
[2-\beta^{-1}(t)] {\bf l}_i'(t)+{\textstyle \frac{h}{2}} {\bf k}_i(t)
\end{eqnarray}
which satisfy the interpolations ${\bf v}_i'(t)=\frac12 [{\bf v}_i(t-
{\textstyle \frac{h}{2}})+{\bf v}_i'(t+{\textstyle \frac{h}{2}})]$, ${\bf
l}_i'(t)=\frac12 [{\bf l}_i(t-{\textstyle \frac{h}{2}})+{\bf l}_i'(t+
{\textstyle \frac{h}{2}})]$ and the constant-temperature condition
$T(\{{\bf v}_i'(t), {\bf \Omega}'_i(t) \})=T_0$, where ${\bf \Omega}'_i
(t)={\bf J}_i^{-1} {\bf A}_i(t) {\bf l}'_i(t)$. Finally, the translational
and orientational positions are updated according to the same equations
replacing ${\bf v}_i(t+{\textstyle \frac{h}{2}})$ by ${\bf v}_i'(t+
{\textstyle \frac{h}{2}})$ and ${\bf l}_i(t+{\textstyle \frac{h}{2}})$
by ${\bf l}_i'(t+{\textstyle \frac{h}{2}})$.

\subsection{Revised leapfrog algorithm}

As has been established [29, 30], the rotational leapfrog algorithm,
described in the preceding subsection, exhibits rather poor long-term
stability of energy with respect to atomic-constraint integrators [4--7],
for example. Moreover, it requires iterative solutions and does not conserve
the unit norm and orthonormality of quaternions and orientational matrices.
For this reason, a question arises how about the existence of a revised
scheme which is free of all these drawbacks and which has all advantages
of the standard approach. We shall show now that such a scheme really
exists.

First of all, one points out some factors which can explain bad stability
properties of the standard scheme. When calculating orientational variables,
the Fincham's algorithm uses up three additional estimators, namely, the
propagations for on-step angular momentum ${\bf l}_i(t)$ (Eq.~(12)) and
mid-step time derivative ${\bf {\dot S}}_i(t+{\textstyle \frac{h}{2}})$
(Eq.~(14)) as well as the prediction (Eq.~(16)) of angular momentum ${\bf
{\tilde l}}_i(t+h)$. Among these only the first two evaluations can be
classified as interpolations which correspond to a simple averaging over
the two nearest neighbouring values. At the same time, the last prediction
(16) presents, in fact, an extrapolation that is, strictly speaking,
beyond the leapfrog framework. Indeed, applying equation (12) for the next
step time $t \equiv t+h$ yields the following interpolated values ${\bf
l}_i(t+h)={\bf l}_i(t+{\textstyle \frac{h}{2}}) + {\textstyle \frac{h}{2}}
{\bf k}_i(t+h)$ for angular momenta, which differ from previously predicted
ones, i.e., ${\bf {\tilde l}}_i(t+h) \ne {\bf l}_i(t+h)$ and, as a
consequence, ${\bf {\tilde \Omega}}_i(t+h) \ne {\bf \Omega}_i(t+h)$.
Extrapolations are commonly used in low-precision explicit schemes and
they should be absent in more accurate implicit integrators.

The main idea of the revised approach is to derive an implicit equation
for ${\bf S}_i(t+h)$ reducing the number of auxiliary interpolations to
a minimum and involving no extrapolations. This can be realized starting
from the same evaluation (10) for mid-step angular momenta, but treating
the time derivatives ${\bf {\dot S}}_i(t+{\textstyle \frac{h}{2}})={\bf H}
({\bf \Omega}_i(t+{\textstyle \frac{h}{2}})) {\bf S}_i(t+{\textstyle \frac{h}
{2}})$ in a somewhat other way. As was mentioned earlier, these derivatives
are necessary to evaluate orientational positions (Eq.~(13)), and they
require the knowledge of two per body quantities, namely, ${\bf \Omega}_i
(t+{\textstyle \frac{h}{2}})$ and ${\bf S}_i(t+{\textstyle \frac{h}{2}})$.
It is crucial to remark that since the advanced angular momenta ${\bf l}_i
(t+{\textstyle \frac{h}{2}})$ are already known, such two quantities are
not independent but connected between themselves by the relation
\begin{equation}
{\bf \Omega}_i(t+{\textstyle \frac{h}{2}})=
{\bf J}_i^{-1} {\bf A}_i(t+{\textstyle \frac{h}{2}})
{\bf l}_i(t+{\textstyle \frac{h}{2}}) \, .
\end{equation}
Then, as can be seen easily, the calculation of ${\bf {\dot S}}_i(t+
{\textstyle \frac{h}{2}})$ is reduced to a propagation of one variable
only, namely, ${\bf S}_i(t+{\textstyle \frac{h}{2}})$. It is quite
naturally to perform this propagation by writing
\begin{equation}
{\bf S}_i(t+{\textstyle \frac{h}{2}})=
\frac12 \Big[{\bf S}_i(t)+{\bf S}_i(t+h) \Big] + {\cal O}(h^2)
\end{equation}
and the algorithm proceeds as follows
\begin{equation}
{\bf S}_i(t+h)={\bf S}_i(t) + h \,
{\bf H}({\bf \Omega}_i(t+{\textstyle \frac{h}{2}})) \,
{\bf S}_i(t+{\textstyle \frac{h}{2}}) + {\cal O}(h^3) \, .
\end{equation}

Taking into account expressions (19) and (20), matrix equation (21)
constitutes an implicit system for unknown elements of ${\bf S}_i(t+h)$.
As for the usual scheme, the system can be solved iteratively, putting
initially ${\bf S}_i^{(0)}(t+h)$ for ${\bf S}_i(t+h)$ in all nonlinear
terms collected in the right-hand side of (21). Then the obtained values
for ${\bf S}_i(t+h)$ in the left-hand side are considered as initial
guesses for the next iteration. The convergence of iterations is
justified by the smallness of nonlinear terms which are proportional
to the step size $h$.

In such a way, we have derived a new leapfrog algorithm to integrate
orientational degrees of freedom. It involves only one auxiliary
interpolation (20) which is completely in the spirit of the leapfrog
framework. Moreover, this interpolation concerns the most slow variables
${\bf S}_i$, rather than their more fast time derivatives ${\bf {\dot
S}}_i$ and angular momenta ${\bf l}_i$, thus, leading to an increased
precision of the calculations. When on-step temperature $T(t)$ is required,
for instance to check the energy conservation, we can apply usual
interpolation (12) of angular momenta and relation (11) for velocities.
It is worth underlining that, unlike the standard rotational integrator,
the angular-momentum interpolation errors are not introduced into
trajectories (21) produced by the revised algorithm at least within
the energy-conserving dynamics.

The extension of the revised scheme to a thermostatted version is trivial.
Using the calculated temperature $T(t)$ we define the scaling factor
$\beta(t)=\sqrt{T_0/T(t)}$. The mid-step angular momenta ${\bf l}_i(t+
{\textstyle \frac{h}{2}})$ are then replaced by their modified values
${\bf l}'_i(t+{\textstyle \frac{h}{2}})$ (see Eq.~(18)) and substituted
into Eq.~(19) to continue the integration process according to equations
(20) and (21).

Besides the evident simplicity of the revised approach with respect to the
standard scheme, a very nice surprise is that the unit norm of quaternions
and the orthonormality of orientational matrices appear to be now by
numerical integrals of motion. Indeed, considering the quantity ${\bf
\Omega}_i(t+{\textstyle \frac{h}{2}})$ as a parameter and explicitly
using coordinate interpolation (20), we can present Eq.~(21) in the
equivalent form
\begin{equation}
{\bf S}_i(t+h)=
[{\bf I}-{\textstyle \frac{h}{2}} {\bf H}_i(t+{\textstyle \frac{h}{2}})]^{-1}
[{\bf I}+{\textstyle \frac{h}{2}} {\bf H}_i(t+{\textstyle \frac{h}{2}})]
{\bf S}_i(t) \, ,
\end{equation}
where ${\bf H}_i(t+{\textstyle \frac{h}{2}}) \equiv {\bf H}({\bf \Omega}_i
(t+{\textstyle \frac{h}{2}}))$ and it is understood that ${\bf I}$ designates
either three- or four-dimensional unit matrix in the principal-axis or
quaternion domain, respectively. It can be checked readily that the matrix
$({\bf I}-{\bf \Theta})^{-1} ({\bf I}+{\bf \Theta})$ is orthonormal for an
arbitrary skewsymmetric matrix ${\bf \Theta}^+=-{\bf \Theta}$. As far as
the matrix ${\bf H}$ is skewsymmetrical by definition, the following
important statement emerges immediately. If initially the orthonormality
of ${\bf S}_i(t)$ is fulfilled, it will be satisfied perfectly for the
advanced matrices ${\bf S}_i(t+h)$ as well, despite an approximate character
of the integration process. Thus, no artificial or constraint normalizations
and no recursive procedures are necessary to conserve the rigidness of
molecules.

The alternative presentation (22) may be more useful for iterating since
it provides the orthonormality of ${\bf S}_i(t+h)$ at each iteration step
and leads to an increased speed of the convergence. Because of this, we
show Eq.~(22) more explicitly,
\vspace{9pt}
\begin{eqnarray}
\textstyle
{\bf q}_i(t+h)&=&\frac{{\bf I}\,[1-{\textstyle \frac{h^2}{16}}
\Omega_i^2(t+{\textstyle \frac{h}{2}})]+h
{\bf Q}_i}{1+{\textstyle \frac{h^2}{16}}
\Omega_i^2(t+{\textstyle \frac{h}{2}})}
\, {\bf q}_i(t) \equiv {\bf G}_i(t,h) \, {\bf q}_i(t) \, , \\ [15pt]
{\bf A}_i(t+h)&=&\frac{{\bf I}\,[1-{\textstyle \frac{h^2}{4}}
\Omega_i^2(t+{\textstyle \frac{h}{2}})]+h{\bf W}_i+
{\textstyle \frac{h^2}{2}}{\bf P}_i}{1+
{\textstyle \frac{h^2}{4}} \Omega_i^2(t+{\textstyle \frac{h}{2}})} \,
{\bf A}_i(t) \equiv {\bf D}_i(t,h) \, {\bf A}_i(t) \, ,
\end{eqnarray}

\vspace{9pt}

\noindent
for the cases of quaternion and entire-rotation-matrix representations,
respectively, where expressions (3) and (5) for matrices ${\bf W}_i \equiv
{\bf W}({\bf \Omega}_i(t+{\textstyle \frac{h}{2}}))$ and ${\bf Q}_i \equiv
{\bf Q}({\bf \Omega}_i(t+{\textstyle \frac{h}{2}}))$ have been taken into
account, ${\bf G}_i(t,h)$ and ${\bf D}_i(t,h)$ are orthonormal evolution
matrices, and $[{\bf P}_i]_{\alpha \beta}={\Omega}_\alpha^i {\Omega}_\beta^i$
denotes a symmetric matrix which, like ${\bf W}_i$ and ${\bf Q}_i$, is
calculated using principal angular velocities (19). In view of the equalities
${\bf W}_i^2={\bf P}_i-\Omega_i^2 {\bf I}$ and ${\bf Q}_i^2=-\frac14
\Omega_i^2 {\bf I}$, the evolution matrices can be cast also in the
matrix-exponential forms
\begin{eqnarray}
{\bf G}_i(t,h)&=&{\rm \bf exp}[\phi_i {\bf Q}_i/
\Omega_i]_{t+\frac{h}{2}} \, , \ \ \ \ \ \
\phi_i=2 \arcsin \frac{\frac{h}{2} \Omega_i(t+{\textstyle \frac{h}{2}})}
{1+{\textstyle \frac{h^2}{16}} \Omega_i^2(t+{\textstyle \frac{h}{2}})} \, ,
\nonumber \\ [-2pt] \\ [-2pt]
{\bf D}_i(t,h)&=&{\rm \bf exp}[\varphi_i {\bf W}_i/
\Omega_i]_{t+\frac{h}{2}} \, , \ \ \ \ \
\varphi_i=\arcsin \frac{h \Omega_i(t+{\textstyle \frac{h}{2}})}
{1+{\textstyle \frac{h^2}{4}} \Omega_i^2(t+{\textstyle \frac{h}{2}})} \, .
\nonumber
\end{eqnarray}

\vspace{9pt}

\noindent
Then it becomes clear that the matrices ${\bf D}_i$ and ${\bf G}_i$ define
three- and four-dimensional rotations on angles $\varphi_i$ and $\phi_i$
in the laboratory frame and quaternion space, respectively. In the first
case the rotation is carried out around the unit vector ${\bf \Omega}_i/
\Omega_i|_{t+\frac{h}{2}}$, whereas in the second one it is performed
around an orth which is perpendicular to all four orths of quaternion
space.

\subsection{Avoidance of iterative solutions}

Another excellent feature of the algorithm is that within the entire matrix
representation, equation (24) can be handled in a non-iterative way using
so-called quasianalytical solutions for mid-step angular velocities ${\bf
\Omega}_i(t+{\textstyle \frac{h}{2}})$. To show this we first perform a
set of further transformations. Remembering that now ${\bf S}_i \equiv {\bf
A}_i$ and ${\bf H}_i \equiv {\bf W}_i$, one adds the matrix ${\bf A}_i(t)$
to the both sides of Eq.~(21) and divides the obtained equation by factor 2.
Then using coordinate propagation (20) leads to
\begin{equation}
{\bf A}_i(t+{\textstyle \frac{h}{2}})={\bf A}_i(t) + {\textstyle \frac{h}{2}}
{\bf W}({\bf \Omega}_i(t+{\textstyle \frac{h}{2}})) \,
{\bf A}_i(t+{\textstyle \frac{h}{2}}) \, .
\end{equation}
Multiplying Eq.~(26) on the matrix ${\bf J}_i^{-1}$ from the left and
additionally on the vector ${\bf l}_i(t+{\textstyle \frac{h}{2}})$ from
the right, and taking into account definition (19) yields
\begin{equation}
{\bf \Omega}_i(t+{\textstyle \frac{h}{2}})=
{\bf J}_i^{-1} {\bf A}_i(t) {\bf l}_i(t+{\textstyle \frac{h}{2}}) +
{\textstyle \frac{h}{2}} {\bf J}_i^{-1}
{\bf W}({\bf \Omega}_i(t+{\textstyle \frac{h}{2}})) {\bf J}_i
{\bf \Omega}_i(t+{\textstyle \frac{h}{2}}) \, .
\end{equation}

Therefore, the iterative problem is much simplified, because it is reduced
to finding solutions to three-dimensional vector equation (27) for three
unknown components $\Omega_X$, $\Omega_Y$ and $\Omega_Z$ of ${\bf \Omega}_i
(t+{\textstyle \frac{h}{2}})$ rather than to matrix equation (24) (or (21))
for nine unknowns elements of ${\bf A}_i(t+h)$. Equation (27) can be solved
iteratively again, choosing ${\bf J}_i^{-1} {\bf A}_i(t) {\bf l}_i(t+
{\textstyle \frac{h}{2}})$ as the initial guess for ${\bf \Omega}_i(t+
{\textstyle \frac{h}{2}})$.

A next simplification lies in the following. Let us rewrite equation (27)
in the explicit form
\begin{eqnarray}
\Omega_X&=&\theta_X+h \varrho_X \Omega_Y \Omega_Z \, , \nonumber \\
\Omega_Y&=&\theta_Y+h \varrho_Y \Omega_Z \Omega_X \, , \\
\Omega_Z&=&\theta_Z+h \varrho_Z \Omega_X \Omega_Y \, , \nonumber
\end{eqnarray}
where $\varrho_X=(J_{YY}^i-J_{ZZ}^i)/(2J_{XX}^i)$, $\varrho_Y=(J_{ZZ}^i-
J_{XX}^i)/(2J_{YY}^i)$, $\varrho_Z=(J_{XX}^i-J_{YY}^i)/(2J_{ZZ}^i)=-
(\varrho_X+\varrho_Y)$, and $\theta_{X,Y,Z}$ are the components of known
vector ${\bf J}_i^{-1} {\bf A}_i(t) {\bf l}_i(t+{\textstyle \frac{h}{2}})$,
keeping in mind that vector ${\bf l}_i(t+{\textstyle \frac{h}{2}})$ must
be replaced by ${\bf l}'_i(t+{\textstyle \frac{h}{2}})$ in the case of
temperature-conserving dynamics. Unless
$J_{XX}^i \ne J_{YY}^i \ne J_{ZZ}^i$, the system of equations (28) appears
to be linear and, therefore, it can easily be solved exactly (see subsect.
III. D, where specific models are described). Here, we consider the most
general case when all the principal moments of molecules are different and
assume for definiteness that $J_{XX}^i < J_{YY}^i < J_{ZZ}^i$. Then the
first two unknowns $\Omega_X$ and $\Omega_Y$ are the most fast variables
and they should be excluded from the iteration to increase the convergence.
Such an excluding indeed can be realized solving the first two equations
of (28) with respect to $\Omega_X$ and $\Omega_Y$. The result is
\begin{equation}
\Omega_X=\frac{\theta_X + h \varrho_X \theta_Y \Omega_Z}
{1+h^2 \nu^2 {\Omega_Z}^2} \, , \ \ \ \ \
\Omega_Y=\frac{\theta_Y + h \varrho_Y \theta_X \Omega_Z}
{1+h^2 \nu^2 {\Omega_Z}^2} \, ,
\end{equation}
where $0 < \nu^2=- \varrho_X \varrho_Y \le 1/4$. The last inequalities
follow from the requirements $J_{\alpha\alpha} > 0$ and $J_{\alpha\alpha}
\le J_{\beta\beta} + J_{\gamma\gamma}$ imposed on principal moments of
inertia, where $(\alpha,\beta,\gamma)$ denote an array of three cyclic
permutations of $(X,Y,Z)$. In view of (29), only the third equation
of system (28) really needs to be iterated with respect to one variable
$\Omega_Z$. Since $\Omega_Z$ is the most slow quantity, a well convergence
can be guaranteed even for not so well normally behaved case as an almost
linear body, when $J_{XX}^i \ll J_{YY}^i < J_{ZZ}^i$.

Finally, one considers the question of how to obviate iterative solutions
at all. Substituting the result (29) into the third equation of system (28)
and presenting the $Z$th component of the angular velocity in the form
$\Omega_Z=s_0+\delta$ yields the following algebraic equation
\begin{equation}
a_0+a_1 \delta+a_2 \delta^2+a_3 \delta^3+a_4 \delta^4+a_5 \delta^5=0
\end{equation}
with the coefficients
\begin{eqnarray}
a_0&=&(s_0 - \theta_Z) \vartheta_+^2 -
h \varrho_Z [ \theta_X \theta_Y \vartheta_-
+ h (\varrho_Y \theta_X^2 + \varrho_X \theta_Y^2) s_0 ] \, , \nonumber \\
a_1&=&\vartheta_+ - h^2 \{ (\varrho_Y \theta_X^2 + \varrho_X \theta_Y^2)
\varrho_Z - \nu^2 s_0 [(5 s_0 - 4 \theta_Z) \vartheta_+
+ 2 h \theta_X \theta_Y  \varrho_Z ] \} \, ,\nonumber \\
a_2&=&h^2 \nu^2 [6 s_0-2 \theta_Z + h \varrho_Z \theta_X \theta_Y
+ h^2 \nu^2 s_0^2 (10 s_0-6 \theta_Z) ] \, , \\
a_3&=& 2 h^2 \nu^2 [1 + h^2 \nu^2 s_0 (5 s_0 - 2 \theta_Z)] \, ,\nonumber \\
a_4&=&h^4 \nu^4 (5 s_0 - \theta_Z) \, , \ \ a_5 = h^4 \nu^4 \, , \nonumber
\end{eqnarray}
where $\vartheta_{\pm}=1 \pm h^2 \nu^2 s_0^2$. The equation (30) is fifth
order and the corresponding solutions for $\Omega_Z^i$ are independent on
parameter $s_0$, provided the unknown $\delta$ is precisely determined.
However, as is well known, only algebraic equations of fourth or less order
allow to be solved in quadratures.

The main idea of quasianalytical solutions consists in the fact that actual
MD simulations are performed with relative small values of the time step
$h$. Then it is necessary to choose the parameter $s_0$ as a good prediction
for $\Omega_Z$ to be entitled to ignore high-order terms in the left-hand
side of Eq.~(30). The simplest choice for this can be found putting
$\Omega_X \Omega_Y = \theta_X \theta_Y+{\cal O}(h)$ in the right-hand side
of the third equation of system (28). As a result, one obtains
\begin{equation}
s_0=\theta_Z+h\varrho_Z \theta_X \theta_Y
\end{equation}
that represents the original value of $\Omega_Z$ with second-order truncation
errors, so that $\delta={\cal O}(h^2)$. It is easy to see that in this case
the two last terms $a_4 \delta^4$ and $a_5 \delta^5$ in the left-hand side
of Eq.~(30) behaves as ${\cal O}(h^{12})$ and ${\cal O}(h^{14})$,
respectively. Taking into account the smallness of $h$, such terms can
merely be omitted without any loss of the precision, because they involve
uncertainties of order ${\cal O}(h^{12})$ into the solutions and appear
to be too small with respect to third-order truncation errors
${\cal O}(h^3)$ inherent initially in the algorithm.

Eq.~(30) is now transformed to the third-order algebraic equation
\begin{equation}
a_0+a_1 \delta+a_2 \delta^2+a_3 \delta^3={\cal O}(h^{12})
\end{equation}
which can easily be solved analytically. The result is
\begin{eqnarray}
\delta_1&=&-{\textstyle \frac13} a_2/a_3+c-b/c +{\cal O}(h^{12}) \, , \\
\delta_{2,3}&=&-{\textstyle \frac13} a_2/a_3-{\textstyle \frac12}
[c-b/c \pm{\rm i} \sqrt{3} (c+b/c)]+{\cal O}(h^{12}) \, , \nonumber
\end{eqnarray}
where
\vspace{-6pt}
\begin{eqnarray}
b&=&{\textstyle \frac19}(3 a_1 a_3 - a_2^2)/a_3^2 \, , \nonumber \\
c&=&(p + \sqrt{b^3+p^2})^{1/3} , \\  p&=&{\textstyle \frac{1}{54}}
(9 a_1 a_2 a_3 - 27 a_0 a_3^2 - 2 a_2^3)/a_3^3 \nonumber \, .
\end{eqnarray}
Among three solutions (34), only the first one $\delta_1$ is real and
satisfies the physical boundary condition $\sim h^2$ when $h$ goes
to zero (the other two solutions $\delta_{2,3}$ are purely imaginary
at $h \to 0$ and they tend to infinity as $\sim \pm {\rm i}/h$).

Therefore, the desired $Z$th component of the angular velocity is
\begin{equation}
\Omega_Z=s_0+\delta_1 \, .
\end{equation}
The rest two components $\Omega_X$ and $\Omega_Y$ are reproduced on the
basis of equalities (29). The obtained vector ${\bf \Omega}_i(t+{\textstyle
\frac{h}{2}}) \equiv (\Omega_X, \Omega_Y, \Omega_Z)$ is substituted into
equation (24) to perform the explicit evaluation for advanced orientational
matrices ${\bf A}_i(t+h)$. This completes the algorithm.

\subsection{Implementations for particular models}

There are two main classes of models for interacting rigid bodies, which
most frequently are applied in MD simulations. More realistic so-called
interaction site models can be related to the first class. For these
models, each $i$th body of the system is considered as a molecule which
in its turn is composed of $M_i$ point interaction sites (atoms). The
rigid structure of molecules is completely defined by time-independent
vector-positions ${\bf \Delta}_i^a$ ($a=1,\ldots,M_i$) of atom $a$ within
molecule $i$ in the body frame, whereas these positions in the laboratory
frame are: ${\bf r}_i^a(t)={\bf r}_i(t)+{\bf A}_i^+(t) {\bf \Delta}_i^a$.
Using the known site-site potentials $u_{ij}^{ab}$, the desired molecular
forces and torques can easily be computed as ${\bf f}_i=\sum_{j(j \ne
i);a,b}^{N;M_i,M_j} {\bf f}_{ij}^{ab}(|{\bf r}_i^a-{\bf r}_j^b|)$ and
${\bf k}_i=\sum_{j(j \ne i);a,b}^{N;M_i,M_j} ({\bf r}_i^a-{\bf r}_i)
{\mbox{\boldmath $\times$}} {\bf f}_{ij}^{ab}$, respectively, where ${\bf
f}_{ij}^{ab}=-\partial u_{ij}^{ab}/\partial {\bf r}_{ij}^{ab}$ and ${\bf
r}_{ij}^{ab}={\bf r}_i^a-{\bf r}_j^b$. The second class is point molecules
($\max_{i;a,b} |{\bf \Delta}_i^a-{\bf \Delta}_i^b| \to 0$) with embedded
multipoles. The most popular model belonging this class is a system of point
electro-dipoles. The molecular forces and torques caused by dipole-dipole
interactions can be calculated using the relations ${\bf f}_i=\sum_{j(j \ne
i)}^N \frac{3}{r_{ij}^5} [{\bf r}_{ij} \{ {\mbox{\boldmath $\mu$}}_i {\mbox
{\boldmath $\cdot$}} {\mbox{\boldmath $\mu$}}_j-\frac{5}{r_{ij}^2} ({\mbox
{\boldmath $\mu$}}_i {\mbox{\boldmath $\cdot$}} {\bf r}_{ij}) ({\mbox
{\boldmath $\mu$}}_j {\mbox{\boldmath $\cdot$}} {\bf r}_{ij}) \} + {\mbox
{\boldmath $\mu$}}_i ({\mbox{\boldmath $\mu$}}_j {\mbox{\boldmath $\cdot$}}
{\bf r}_{ij}) + {\mbox{\boldmath $\mu$}}_j ({\mbox{\boldmath $\mu$}}_i
{\mbox{\boldmath $\cdot$}} {\bf r}_{ij})]$ and ${\bf k}_i=\sum_{j(j \ne i)}^N
\frac{1}{r_{ij}^3} {\mbox{\boldmath $\mu$}}_i {\mbox{\boldmath $\times$}}
[\frac{3}{r_{ij}^2} {\bf r}_{ij} ({\mbox{\boldmath $\mu$}}_j {\mbox{\boldmath
$\cdot$}} {\bf r}_{ij}) - {\mbox{\boldmath $\mu$}}_j]$, respectively, where
${\bf r}_{ij}={\bf r}_i-{\bf r}_j$ and ${\mbox{\boldmath $\mu$}}_i$ denotes
the dipole moment of $i$th molecule.

Although the proposed algorithm can be implemented for arbitrary rigid
models, some simplifications with respect to the general formulation are
possible using special properties of the body. The simplest case is bodies
with a spherical distribution of mass, when all the moments of inertia are
equal between themselves, i.e., when $J_{XX}^i=J_{YY}^i=J_{ZZ}^i \equiv
J_i$ and, thus, ${\bf J}_i=J_i{\bf I}={\bf j}_i$. Then it is more convenient
to work with equations (2) for rotational matrices, presented in terms of
angular velocities ${\mbox{\boldmath $\omega$}}_i={\bf j}_i^{-1} {\bf l}_i=
{\bf l}_i/J_i$ in the laboratory frame, i.e., with ${\rm d} {\bf A}_i/{\rm
d} t={\bf A}_i {\bf W}({\mbox {\boldmath $\omega$}}_i)$. The leapfrog
trajectories for these equations are obvious: ${\bf A}_i(t+h)={\bf A}_i(t)
{\rm \bf exp}[\varphi_i {\bf W}_i({\mbox{\boldmath $\omega$}}_i)/\omega_i]_
{t+\frac{h}{2}}$, where ${\mbox{\boldmath $\omega$}}_i(t+{\textstyle \frac
{h}{2}})={\bf l}_i(t+{\textstyle \frac{h}{2}})/J_i$ and $\varphi_i=\arcsin
[h \omega_i/(1+{\textstyle \frac{h^2}{4}} \omega_i^2)]_{t+\frac{h}{2}}$.

For some particular models, the orientational part of the intermolecular
potential can be expressed using only unit three-component vectors
${\mbox{\boldmath $\rho$}}_i$ passing through the centres of mass of
molecules. The examples are point dipole interactions, when ${\mbox
{\boldmath $\rho$}}_i \equiv {\mbox{\boldmath $\mu$}}_i/\mu_i$, or when
all force sites of the molecule are aligned along ${\mbox{\boldmath
$\rho$}}_i$, resulting in torques which are perpendicular to ${\mbox
{\boldmath $\rho$}}_i$, i.e., ${\bf k}_i {\mbox{\boldmath $\cdot$}}
{\mbox{\boldmath $\rho$}}_i=0$. If then additionally ${\bf J}_i=J_i{\bf
I}$ (for the last example this can be possible when forceless mass sites
are placed in such a way to ensure the spherical mass distribution), it
is no longer necessary to deal with orientational matrices or quaternions.
In this case the equation for ${\mbox{\boldmath $\rho$}}_i$ looks as ${\rm
d} {\mbox{\boldmath $\rho$}}_i/{\rm d} t={\bf W}^+({\mbox{\boldmath
$\omega$}}_i) {\mbox{\boldmath $\rho$}}_i$ with the solution ${\mbox
{\boldmath $\rho$}}_i(t+h)={\rm \bf exp}[-\varphi_i {\bf W}_i({\mbox
{\boldmath $\omega$}}_i)/\omega_i]_{t+\frac{h}{2}} {\mbox{\boldmath
$\rho$}}_i(t)$.

For molecules with cylindric distribution of mass sites, when two of three
of principal moments of inertia are equal, the numerical trajectory can also
be determined in a simpler manner. Let us assume for definiteness that
$J_{XX}^i=J_{YY}^i \ne J_{ZZ}^i$ and $J_{ZZ}^i \ne 0$. Then arbitrary two
perpendicular between themselves axes, lying in the plane perpendicular to
the $Z$th principal axis, can be considered initially as $X$- and $Y$-th
principal orths. Since now $\varrho_Z=0$, the $Z$th component of the
angular velocity is found automatically, $\Omega_Z=\theta_Z$. As in the
general case, the two rest solutions $\Omega_X$ and $\Omega_Y$ of system
(28) are calculated on the basis of Eq.~(29) taking into account that
$\varrho_Y=-\varrho_X$, whereas the orientational matrices are evaluated
via Eq.~(24).

A special attention must be paid for purely linear molecules when
$J_{XX}^i=J_{YY}^i \ne J_{ZZ}^i=0$ and each body has two, instead of free,
orientational degrees of freedom. The relative positions ${\bf r}_i^a
(t)-{\bf r}_i(t)=\Delta_i^a {\mbox{\boldmath $\rho$}}_i(t)$ of all atoms
within a linear molecule can be expressed in terms of an unit vector
${\mbox{\boldmath $\rho$}}_i$ and besides ${\bf k}_i {\mbox{\boldmath
$\cdot$}} {\mbox{\boldmath $\rho$}}_i=0$ one finds that $[{\bf L}_i]_Z=
J_{ZZ}^i {\Omega_Z^i}=0$. The rotational part $\frac12 (J_{XX}^i
{\Omega_X^i}^2+J_{YY}^i {\Omega_Y^i}^2)$ of the kinetic energy is also
indifferent to the $Z$th component $\Omega_Z^i$ of the principal angular
velocity. Such a component causes irrelevant rotations of the molecule
around ${\mbox{\boldmath $\rho$}}_i$-axis and it does not lead to any
change of ${\bf r}_i^a$ and the potential energy. It can be shown that
the angular-momentum approach allows to reproduce the correct time
evolution of two-dimensional unit vector ${\mbox{\boldmath $\rho$}}_i$
by the three-dimensional leapfrog rotation ${\mbox{\boldmath $\rho$}}_i
(t+h)={\rm \bf exp}[\varphi_i {\bf W}_i({\bf \Omega}_i)/\Omega_i]_{t+
\frac{h}{2}} {\mbox{\boldmath $\rho$}}_i(t)$ putting formally $\Omega_Z^i
\equiv 0$, where two other components of ${\bf \Omega}_i(t+{\textstyle
\frac{h}{2}})$ are $\Omega_X=\theta_X$ and $\Omega_Y=\theta_Y$ (this
immediately follows from Eq.~(28)). Planar molecules do not present a
specific case within our approach and they are handled in the usual way
as tree-dimensional bodies.

\section{Numerical verification of the algorithm}

The system chosen for numerical tests was the TIP4P model ($M=4$) of water
[32] at a density of $m N/V=1$ g cm$^{-3}$ and a temperature of $T=298$ K.
Because of the low moments of inertia of the water molecule and the large
torques due to the site-site interactions, such a system should provide
a very severe test for rotational algorithms. In order to reduce cut-off
effects to a minimum we have applied an interaction site reaction field
geometry [33] and a cubic sample with $N=256$ molecules. All runs were
started from an identical well equilibrated configuration. The MD
simulations have been carried out in both energy-conserving (NVE) and
thermostatted (NVT) ensembles. The equations of rotational motion were
integrated using the standard quaternion integrator [27] and our revised
leapfrog algorithm. As far as water is usually [34] simulated in an NVE
ensemble by the atomic-constraint technique [4, 5], the corresponding
calculations on this approach and the angular-velocity Verlet method [28]
were performed for the purpose of comparison as well. All the approaches
required almost the same computer time per step given that near 97\% of
the total time were spent to evaluate pair interactions.

The following thermodynamic quantities were evaluated: total energy,
potential energy, temperature, specific heat at constant volume, and
mean-square forces and torques. The structure of the TIP4P water was
studied by determining the oxygen-oxygen and hydrogen-hydrogen radial
distribution functions (RDFs). Orientational relaxation was investigated
by evaluating the molecular dipole-axis autocorrelations. Centre-of-mass
and angular-velocity time autocorrelation functions were also found. To
reduce statistical noise, the measurements were averaged over 20 000
time steps.

In the case of NVE dynamics to verify whether the phase trajectories are
produced properly, we applied the most important test on conservation of
total energy $E$ of the system. The total energy fluctuations ${\cal
E}=[\langle (E-\langle E \rangle)^2 \rangle]^{1/2}/|\langle E \rangle|$
as functions of the length of the simulations over 10 000 time steps are
plotted in Fig.~1 (a)--(d) at four fixed step sizes, $h=$ 1, 2, 3 and 4
fs. Both principal-axis (the boldest curves in subsets (a)--(d)) and
quaternion representations were used to integrate the equations by the
revised leapfrog algorithm. It has been established that the functions
${\cal E}$ corresponding to these representations are practically the
same. For this reason and to simplify the graph notations the results
obtained within quaternion variables are shown (as crosses) only in
subset (d) of the figure.

As can be seen easily, the standard rotational leapfrog algorithm exhibits
relatively bad stability properties and conserves the energy rather poor
even at the smallest step size considered. It is worth remarking that
investigating the system during shorter time periods with small step sizes,
for example over 1000 time steps with $h=$ 1 fs, one may come to very
misleading conclusions on the energy conservation. A significantly better
pattern is observed for the angular velocity Verlet integrator. However,
the improvements in stability are quite insufficient especially for moderate
and large step sizes ($h \ge$ 3 fs, subsets (c)--(d)). Finally, we can
talk about the best energy conservation and long-term stability for the
atomic-constraint scheme and the revised leapfrog algorithm which lead
to virtually identical results.

It is necessary to emphasize that within the principal-axis representation,
the revised leapfrog trajectories were evaluated using the non-iterative
quasianalytical scheme. The exact solutions (by means of iterations of
Eq.~(28)) were computed too to compare it with quasianalytical values.
No deviation between both trajectories has been found up to $h=10$ fs.
They differed on each step by uncertainties of order round-off errors, so
that the quasianalytical hypothesis appears to be in an excellent accord.
At the same time, the quaternions converged at each step to a relative
tolerance of $10^{-10}$ in average from 6 to 14 iterations with varying
the step size from 2 fs to 6 fs.

We also tried to avoid iterative procedures for quaternion variables by
applying a hybrid leapfrog scheme when the quasianalytical solutions for
mid-step angular velocities are substituted directly into orthogonormal
matrices for quaternion evaluation (23). However, the hybrid scheme leads
to a significant loss of the precision (see the long-dashed curve in subset
(d) of Fig.~1). This is so because equation (27) for angular velocities was
obtained on the basis of interpolation (20) for principal-axis variables,
i.e., when ${\bf A}_i(t+ {\textstyle \frac{h}{2}})=\frac12 [{\bf A}_i(t)+
{\bf A}_i(t+h)]$. Using these velocities in the quaternion space causes an
inconsistency of such an interpolation with the corresponding interpolation
${\bf q}_i(t+ {\textstyle \frac{h}{2}})=\frac12 [{\bf q}_i(t)+{\bf q}_i
(t+h)]$ for quaternions since ${\bf A}_i[{\bf q}_i(t+ {\textstyle \frac{h}
{2}})] \ne \frac12 ({\bf A}_i[{\bf q}_i(t)]+{\bf A}_i[{\bf q}_i(t+h)])$.
Therefore, to follow rigorously the leapfrog framework, the auxiliary
mid-step angular velocities must be involved within the principal-axis
representation exclusively.

No shift of the total energy and temperature was observed during the
revised leapfrog trajectories at $h \le 5$ fs over a length of 20 000
time steps. As is well known, to reproduce features of an NVE ensemble
correctly, the ratio $\Upsilon={\cal E}/{\cal U}$ of the total energy
fluctuations to the corresponding fluctuations ${\cal U}$ of the potential
energy must be within a few per cent. The following levels of ${\cal E}$
at the end of the revised leapfrog trajectories have been obtained: 0.0016,
0.0066, 0.017, 0.030, 0.051 and 0.11 \%. They correspond to $\Upsilon
\approx$ 0.29, 1.2, 3.0, 5.4, 9.1 and 20 \% at $h=$ 1, 2, 3, 4, 5 and 6
fs, respectively, where it was taken into account that ${\cal U} \approx
0.56 \%$ for the investigated system. The deviations in all the rest
measured functions with respect to their benchmark values (obtained in
the atomic-constraint NVE simulations with $h=$ 2 fs) were in a complete
agreement with the corresponding relative deviations $\Upsilon$ in the
total energy conservation. For example, the results of the revised leapfrog
algorithm at $h=$ 2 fs were indistinguishable from the benchmark ones,
whereas they differed as large as around 5\%, 10\% and 20\% with increasing
the time step to 4 fs, 5 fs and 6 fs, respectively. However, the differences
were much smaller than in the case of the standard rotational integrator. We
see, therefore, that step sizes of order 5 fs are still suitable for precise
NVE calculations. Even a time step of 6 fs can be acceptable when a great
precision is not so important, for instance, to achieve an equilibrium state.

What about the NVT simulations? It is well established [27, 35] that
thermostatted versions allow to perform reliable calculations with
significantly greater step sizes than those used within the
energy-conserving dynamics. To confirm such a statement, we have
made NVT runs on the basis of our non-iterative revised leapfrog
algorithm (within principal-axis variables) and a thermostatted
version of the standard implicit integrator [27] of Fincham.

The oxygen-oxygen and hydrogen-hydrogen RDFs, calculated during the revised
leapfrog trajectories for three different step sizes, $h=$ 2, 8 and 10 fs,
are plotted in Fig.~2a by the curves in comparison with the benchmark result
(open circles). Note that the RDFs corresponding to $h=4$ and 6 fs coincide
completely with those for $h=2$ fs and they are not included in the graph.
A similar behaviour of RDFs was identified for the standard rotational
integrator, but the results are somewhat worse especially at $h=$ 8 and 10
fs. No drift of the potential energy was observed at $h \le 10$ fs and
$h < 6$ fs for the revised and standard algorithms, respectively. From the
above, we can conclude that the revised leapfrog integrator is suitable
for simulating even with huge step sizes of 10 fs, because then there is no
detectable difference in RDFs. Other thermodynamic quantities such as the
centre-of-mass and angular-velocity time autocorrelation functions appeared
to be also close to genuine values. Quite recently, it was shown [36] that
the time interval of 10 fs should be considered as an upper theoretical
limit for the step size in MD simulations on water. We see, therefore,
that this limit can be achieved in practice using the revised leapfrog
algorithm.

The molecular dipole-axis time autocorrelation function is the most sensitive
quantity with respect to varying the step size. Such a function obtained
within the standard (S) and revised (R) schemes at five fixed step sizes,
$h=$ 2, 4, 6, 8 and 10 fs, is presented in Fig.~2b. For $h\le$ 6 fs the
results of S- and R-schemes are indistinguishable between themselves. With
increasing the step size to 8 fs or higher we can observe a systematic
discrepancy which is smaller in the case of the R-scheme. Reliable results
can be obtained here at time steps of $h<$ 8 fs for both the standard
and revised schemes. However within the standard approach, the solutions
converged too slow already at $h=$ 6 fs and they began to diverge at greater
step sizes. To perform the simulations in this case, special time-consuming
transformations to unsure the convergence have been applied. For the revised
integrator which is free of iterations, the computer time did not depend on
the step size.

\section{Conclusions}

During last years there was a slow progress in the improvement of existing
MD techniques concerning the numerical integration of motion for systems
with interacting rigid bodies. We have attempted to remedy such a situation
by formulating a revised angular-momentum approach within the leapfrog
framework. As a result, a new integrating algorithm has been derived.
The revised approach reduces the number of auxiliary interpolations to
a minimum, applies the interpolations to the most slow variables and
avoids any extrapolations. This has allowed to achieve the following two
significant benefits: (i) all final expressions are evaluated explicitly
without involving any iterative procedures, and (ii) the rigidity of bodies
appears to be a numerical integral of motion. Another positive feature
of the algorithm is its simplicity and universality for the implementation
to arbitrary rigid structures with arbitrary types of interactions.

As has been shown on the basis of actual simulations of water, the proposed
algorithm exhibits very good stability properties and conserves the total
energy in microcanonical simulations with the same precision as the
cumbersome atomic-constraint technique. In the case of temperature-conserving
dynamics, reliable calculations are possible with huge step sizes around 10
fs. Such sizes are very close to the upper theoretical limit and
unaccessible in usual approaches.

\vspace{12pt}

{\bf Acknowledgements.} The author would like to acknowledge financial
support by the President of Ukraine.

\newpage

\vspace*{1cm}

\begin{center}
{\large Figure captions}
\end{center}

Fig. 1. The total energy fluctuations as functions of the length of the
NVE simulations on the TIP4P water, evaluated in various techniques at
four fixed time steps: {\bf (a)} 1 fs, {\bf (b)} 2 fs, {\bf (c)} 3 fs
and {\bf (d)} 4 fs.

\vspace{12pt}

Fig. 2. Oxygen-oxygen (O-O) and hydrogen-hydrogen (H-H) radial distribution
functions {\bf (a)}, and orientational relaxation {\bf (b)}, obtained in NVT
simulations on the TIP4P water using the revised ({\bf (a)}, {\bf (b)}) and
standard ({\bf (b)}) leapfrog algorithms. The results corresponding to the
step sizes $h=$ 2, 8 and 10 fs are plotted by bold solid, short-dashed and
thin solid curves, respectively. Additional long-short dashed and dashed
curves in {\bf (b)} correspond to cases of $h=$ 4 and 6 fs. The sets of
curves related to standard and revised integrators are labelled in {\bf (b)}
as "S" and "R", respectively. The benchmark data are shown as open circles.
Note that the standard- and revised-algorithm curves are indistinguishable
in {\bf (b)} at $h=$ 2, 4 and 6 fs.

\end{document}